\documentclass[%
 reprint,
 superscriptaddress,
 amsmath,amssymb,
 aps,
]{revtex4-2}

\usepackage{subfigure}
\usepackage{amsmath,graphicx,amssymb}
\usepackage{multirow,color}
\usepackage{url}
\usepackage{verbatim}
\usepackage{bm}

\begin{document}

\title{Chaotic measures as an alternative to spectral measures for
analysing turbulent flow}

\author{Richard D.J.G. Ho}
\thanks{Correspondence: richh@uio.no}%
\affiliation{Njord centre, Department of Physics, University of Oslo, 0371 Oslo, Norway}
\affiliation{School of Physics and Astronomy, University of Edinburgh, Edinburgh EH9 3FD, United Kingdom}


\author{Daniel Clark}
\affiliation{Thornton Tomasetti , 2 St Davids Drive, KY11 9PF}
\affiliation{School of Physics and Astronomy, University of Edinburgh, Edinburgh EH9 3FD, United Kingdom}

\author{Arjun Berera}
\affiliation{School of Physics and Astronomy, University of Edinburgh, Edinburgh EH9 3FD, United Kingdom}

\date{\today}

\begin{abstract}Turbulence has associated chaotic features. 
In the past couple of decades there has been growing interest in the study of these features as an alternative means of understanding turbulent systems.
Our own input to this effort has been in contributing to
the initial studies of chaos in Eulerian flow using direct
numerical simulation (DNS). In this review we discuss
the progress achieved in the turbulence community
in understanding chaotic measures including our own work.
A central relation between
turbulence and chaos is one by Ruelle
that connects the maximum Lyapunov exponent and the Reynolds number.
The first DNS studies, ours amongst them, in obtaining
this relation has shown the viability of chaotic simulation studies of
Eulerian flow.  Such chaotic measures and associated
simulation methodology provides an
alternative means to probe turbulent flow.
Building on this, we have analyzed the finite time
Lyapunov exponent (FTLE) and studied its fluctuations, and 
found that chaotic measures could be quantified accurately even at small simulation box sizes
where for comparative sizes spectral measures would be inconclusive.
We further highlight
applications of chaotic measures in analyzing phase transition
behavior in turbulent flow and two dimensional thin layer turbulent systems.
This work has shown chaotic measures are an excellent
tool that can be used alongside spectral measures in studying
turbulent flow.
\end{abstract}

\maketitle

\section{Introduction}

Turbulence is a descriptor for fluid flow with highly complex structure, where the velocity varies rapidly with space and time \cite{mccomb1990physics}. Although the flow is
seemingly random, its statistical quantities and their dependence on length scales are reproducible.
These statistics strongly depend on the boundaries and other macroscopic restrictions.
Along with global statistics such as dissipation and turbulent energy, structure functions can characterise turbulent flows. For homogeneous isotropic turbulence (HIT), the structure functions can be related to the spectral quantities, which are statistics that depend on the wave number \cite{mccomb2014homogeneous}.
These are relevant in statistical closure approximations, where higher order structure functions are related to lower order ones \cite{kraichnan1959structure,zhou2021turbulence}.

However, chaotic measures are another set of statistics which are comparatively less well established in the turbulence literature even though there are several advantages of using them over spectral measures. We believe that analysis of the chaotic properties of turbulence presents an untapped potential for a better understanding of turbulence.
Recently new advances have been made in quantifying chaotic measures in turbulence and this review will provide context and framework for these studies, track the evolution of the main ideas, as well as summarise the key implications of them.

In discussing the advantages of chaotic measures, we will very briefly discuss the disadvantages of spectral measures in the context of direct numerical simulation (DNS).
Within the K41 theory, there is the large length scale where energy is inputted and then the dissipation at small length scale.
These scales are decoupled and between them should exist an inertial range where energy transfer occurs from larger to smaller scales.
Prediction of the wave number scaling of this range as well as how the longitudinal structure functions relate to the dissipation can then be made \cite{taylor1938spectrum,kolmogorov1941local,kolmogorov1962refinement}.
Intermittency effects can be said to account for the degree that these predictions agree or do not agree with simulations and experiment \cite{frisch1995turbulence}.
However, with DNS it is often hard to generate a well defined inertial range.
To do so, we need a large separation of scales, which requires a large simulation and thus large computational resources.
Unlike the scaling of any inertial range, many chaotic quantities are more like the turbulent energy or the local dissipation rate - they exist regardless of the existence of a well defined inertial range. 
They thus can be calculated with much more modest simulation sizes.
Laws for scaling such as $S_n(r) \sim r^{\zeta_n}$ for the structure functions, where $\zeta_n$ is the scaling exponent,
require very large Reynolds numbers and a well defined inertial range as shown by Qian \cite{qian1997inertial,qian1999slow,sagaut2008homogeneous}. The fact that these conditions are not required for the chaotic measures adds a strong argument for their use.
It is notable that, even in 1986 as shown by Deissler \cite{deissler1986navier}, robust chaos was observed where an inertial range would definitely not have been found.

\subsection{Advantages of chaotic measures applied to turbulence}

Chaos captures the idea that a small initial difference between two systems
can grow exponentially in time.
We can study an evolving fluid in two ways,
either we may imagine following one fluid parcel in the flow and see how it evolves 
its position in time, which is the Lagrangian approach. 
Alternatively, we may also imagine standing in one position and seeing what the fluid does at that position, which is the Eulerian approach.
These approaches are inextricably linked. 
If we knew the precise details of the flow at every point at all times,
we could determine how each fluid parcel behaved. 
Similarly, if we knew the behaviour of each parcel, 
we could recreate the evolution at each position.

Turbulence displays chaotic dynamics
\cite{bohr1998dynamical} and 
ideas from chaos theory find many different applications in
turbulence including the dispersion of pairs of particles \cite{taylor1922diffusion,richardson1926atmospheric,salazar2009two,biferale2005lagrangian}, the presence of Lagrangian coherent structures 
\cite{haller2015lagrangian}, turbulent mixing \cite{ottino1990mixing}, turbulent transitions 
\cite{eckhardt2007turbulence} and predictability
\cite{lorenz1963deterministic,lorenz1969predictability,aurell1997predictability,leith1971atmospheric,leith1972predictability,boffetta2002predictability}.
Whilst the Lagrangian aspects of chaos in turbulence
have been heavily studied, the Eulerian picture has been comparatively neglected up until the past few years when advances in computing power and theory have allowed novel insights to be found in long pondered questions.
These recent developments are the main focus of this review, including the contributions we have made
to the overall effort.

One of the reasons we may wish to apply chaotic measures to turbulence is in the name of predictability.
We may believe that given more and more accurate information of a system and its governing equations, we will be able to predict the future of that system with more and more precision.
However, chaos means that a small difference from the true evolution will diverge exponentially.
In many short introductions of turbulence, the fact that turbulence is chaotic is mentioned and then not further explored. 
Sometimes a calculation is made of the predictability time, whereby, because the chaos is very high, the predictability is declared to be very low.
This, however, goes against our experience of the atmosphere, which is highly turbulent and chaotic, yet weather prediction is accurate for a non-trivial time frame.

Another reason to apply chaotic measures is to understand the nature of turbulence itself.
Different hypotheses of the nature of turbulence lead to different predictions of the implications for chaos in turbulent dynamics.
These implications can be tested and hypotheses strengthened or discarded.
Beyond theoretical considerations, there are practical ones as well.
Turbulence is frequently studied computationally.
In this review, one of the themes we highlight is that chaotic measures can be used to quantify the resolution of simulations.
In structured flows, this allows us to determine where a simulation needs more resources, thus allowing more efficient computation.
Whilst this review is mainly concerned with chaotic measures in HIT, which is an idealized
case of primarily theoretical interest, it
acts as a testbed to develop the method for
more practical use.  For example, the application of chaotic measures could perhaps be generalised to wave turbulence, which describes a broad category of weakly nonlinear dispersive waves \cite{nazarenko2011wave}, including gravity-capillary wave turbulence \cite{korotkevich2023inverse,falcon2022experiments,pan2014direct}, magneto-hydrodynamic free-surface wave turbulence \cite{kochurin2022three,ricard2023transition}, and acoustic turbulence \cite{griffin2022energy,kochurin2022direct}. Researchers in this field may find the ideas presented in this review as relevant to their own field of study.

\section{Chaotic measures of turbulence}

\subsection{Lyapunov exponents}

The nature of the chaos in a system can be quantified by its Lyapunov spectrum. 
We can define a dynamical system as an $N$-dimensional system of first-order differential equations specifying the time evolution, given by the trajectory $x(t)$, of an initial condition $x(0)$. For an explicit example, consider
\begin{equation}
\frac{{dx(t)}}{{dt}} = F[x(t)], \nonumber
\end{equation}
such as the Navier-Stokes equations.
For an infinitesimally small perturbation to $x(0)$ given by $\delta(0)$, then at later times we have
\begin{equation}
x'_0 (t) = x(t) + \delta(t) \nonumber
\end{equation}
where $\delta(t)$ gives the separation between the two trajectories. 
The maximum Lyapunov exponent, $\lambda_1$, has the greatest influence on the predictability of the system.
It gives a relation for the separation between trajectories $\delta(t)$ for a finite time interval
\begin{equation}
|\delta(t)| \approx |\delta(0)|e^{\lambda_1 t} \ . \nonumber
\end{equation}
Therefore, if $\lambda_1$ is positive, we have an exponential divergence between $x(t)$ and $x_0 (t)$, and the system is said to exhibit deterministic chaos. This rapid divergence between two initially close states represents an extreme dependence on initial conditions.

As long as $\delta(t) = x_0 (t) - x(t)$ is small, it can be considered a tangent vector.
We can then find the time evolution of $\delta(t)$ from the linearized equation
\begin{equation}
\frac{d\delta_i (t)}{dt} = \sum_{j=1}^{d} \frac{\partial F_i}{\partial x_j} \Bigg|_{x(t)} \delta_j (t) \ . \nonumber
\end{equation}
Then there exists a basis ${e_i}$ in this tangent space such that for large $t$ \cite{oseledec1968multiplicative},
\begin{equation}
\delta(t) = \sum_{i=1}^{N} c_i e^{\lambda_i t} e_i. \nonumber
\end{equation}
The ${\lambda_i}$ are known as the Lyapunov exponents of the system and are ordered by convention as $\lambda_1 > \lambda_2 > \dots > \lambda_N$. This collection of exponents is called the Lyapunov spectrum of the system \cite{wolf1985determining}.
In a conservative system, the sum of Lyapunov exponents must be zero \cite{dettmann1996proof}.
As some trajectories diverge exponentially 
(they have positive Lyapunov exponent) others will converge.
A dissipative system with at least one $\lambda_i > 0$ is said to be chaotic.
We will look at this spectrum further in the context of the Kolmogorov-Sinai entropy below.

Under strict mathematical requirements, the exact Lyapunov spectrum requires an infinite amount of time to calculate, which is infeasible.
For estimating the maximum Lyapunov exponent, one can adopt the naive method, where a random small perturbation is made, and the divergence tracked, with a fit made to the subsequent exponential growth.
However, this takes a long time for each Lyapunov exponent calculated and could introduce biases depending on the form of the perturbation.
As such, different generalisations have been made to calculate the maximum Lyapunov exponent.
Two commonly used methods are the finite size and finite time Lyapunov
exponents, denoted FSLE and FTLE respectively.

For a perturbation of an initial size, $\delta$, one may wish to know, on average, how long it takes to grow to a certain tolerance, $\Delta$.
This time is denoted $T(\delta, \Delta)$. We can then define the finite size Lyapunov exponent (FSLE) as \cite{aurell1997predictability}
\begin{equation}
\label{eq:FSLEdef}
\lambda (\delta, \Delta) = \bigg\langle \frac{1}{T(\delta,\Delta)} \bigg\rangle
\text{ln} \bigg( \frac{\Delta}{\delta} \bigg) \ ,
\end{equation}
where the angled brackets represent an average, and
which recovers the usual definition of the maximum Lyapunov exponent when $\delta$ and
$\Delta$ are infinitesimal.
If disturbances grow at a rate equal to the local eddy turnover time then $\lambda(\delta) \sim \delta^{-2}$ \cite{lorenz1969predictability,boffetta2017chaos}.

We can change the definition of the FSLE in eq.~(\ref{eq:FSLEdef})
so that, instead of fixing an initial tolerance $\Delta$
and then measuring $T (\delta, \Delta)$,
we fix a time $T$ and measure the error $\Delta(\delta, T)$ after that time.
This defines the finite time Lyapunov exponent (FTLE) as
\begin{equation}
\label{eq:FTLEdef}
\lambda (\delta, T) = \frac{1}{T} \text{ln} \bigg( \frac{\langle \Delta(\delta, T) \rangle}{\delta} \bigg) \ ,
\end{equation}
where we will not currently worry about whether this averaging procedure takes
place inside or outside the fraction or the natural logarithm.
We can also define a definition for $\lambda$ which is spatially dependent for
chaos in the Lagrangian sense.
These can be used to characterise the presence of Lagrangian 
coherent structures (LCS) \cite{haller2015lagrangian}.
These are surfaces on which the infinitesimal deformation of the flow is maximised,
represented by extremising surfaces of $\lambda$.

In the Eulerian sense, we can implement the FTLE with the following procedure;
we take the original field $u_1$, and create another field $u_2 = u_1 + \delta u$,
with some randomisation procedure.
Then, we evolve the system over a finite time $T$,
measure the field difference,
and then introduce a successive perturbation, starting the process again.
This subsequent perturbation is given by 
\begin{equation}
    u_2 (x, T) = u_1 (x, T) + \frac{\delta}{\Delta(\delta, T)} \delta u (x,T) \ . \nonumber
\end{equation}
The maximum Lyapunov exponent, which fluctuates in time, $\Tilde{\lambda}$ can then be calculated as
\begin{equation}
    \Tilde{\lambda} = \frac{1}{T} \ln \left(\frac{\Delta}{\delta}\right)  \ , \nonumber
\end{equation}
and this procedure is repeated, obtaining a set of values for $\Tilde{\lambda}$.

One advantage of the FTLE method is that it produces a sample of maximum Lyapunov exponents 
that allows us to perform a statistical analysis of the measurements. 
Compared to the naive method of direct measurement of $|\delta(t)|$, using the FTLE method it is possible to get a large number of maximum Lyapunov exponents 
which can be used to analyse the distribution for a given steady state. 
This procedure shows that time fluctuations in the maximum Lyapunov exponent is quite significant,
which is difficult to show using the direct method. 
However, we also lose the ability to directly control the form of the perturbation,
which can give us important physical information about turbulence.

\subsection{KS Entropy}

The unpredictability of deterministically chaotic systems can also be viewed from an information theoretical viewpoint. To this end, the concept of entropy can be extended to such systems through the Kolmogorov-Sinai (KS) entropy which provides a measure of the information production rate and is an invariant property of the system.

To define the KS entropy, which is a generalization of the classical Shannon entropy to dynamical systems, we begin by performing a discretization of the $d$-dimensional state space of the system, \textit{i.e.}, a space where each point represents a unique state of the system. After discretization the space is composed of cells of side length $l$ and thus volume $l^d$. Our space is equipped with an invariant probability density function, $\rho(x)$, hence we can write the probability of the system being in a given cell as
\begin{equation}
p_i = \int_{P_i} \rho(x) , dV_i. \nonumber
\end{equation}
These probabilities could be used to define a Shannon entropy for the system. However, as it stands, the above definition is not independent of how the space is partitioned, which is required of a system invariant such as the KS entropy \cite{bohr1998dynamical,eckmann1985ergodic}. Moreover, the KS entropy provides the rate of information production thus it's definition should include time dependence. 

We now extend our probabilistic definition to consider trajectories through state space. By looking at these trajectories, we are using the Eulerian framework for chaos. As we have a measure of the probability the system will be located in a given cell, we can define the probability of a trajectory comprised of being in cell $i_1$ at $t = \tau$, then $i_2$ at $t = \tau + \delta t$, and so on until at $t = \tau + n \delta t$, the system is in cell $i_n$ which we denote as $p_{i_1 i_2 \ldots i_n}$. Then, still motivated by the Shannon entropy, we define the entropy of such a trajectory as 

\begin{equation}
H_n = - \sum_{i_1, \ldots, i_n} p_{i_1 i_2 \ldots i_n} \log_2 p_{i_1 i_2 \ldots i_n}. \nonumber
\end{equation}

The difference in entropy between a trajectory at step $n$ and the same trajectory at step $n+1$ provides a measure of information production. Then by taking appropriate limits to remove dependence on trajectory length, sampling frequency and how we choose to discretize space, the KS entropy, $h_{KS}$ is defined \cite{schuster2006deterministic}
\begin{equation}
h_{KS} = \lim_{\delta t \to 0} \lim_{l \to 0} \lim_{N \to \infty} \frac{1}{N \tau} \sum_{n = 0}^{N-1} (H_{n+1} - H_n) \ . \nonumber
\end{equation} Systems which exhibit deterministic chaos have positive, but finite, KS entropy whilst purely random systems have infinite entropy.

A more practical definition of the KS entropy can be found by making a connection with the Lyapunov spectrum of the system through Pesin's theorem \begin{equation}
h_{KS} = \sum_{\lambda_i > 0} \lambda_i \ . \nonumber
\end{equation} Considering once more a discretized state space, if two trajectories are initially located inside the same cell then they are effectively identical. If the system exhibits deterministic chaos then these trajectories will diverge over time, with the rate of divergence along a given eigenvector of the system given by the corresponding Lyapunov exponent. Eventually, this divergence will lead to the two trajectories occupying different cells and thus becoming distinguishable, which can be viewed as producing information. The above definition of the KS entropy in terms of the sum of positive Lyapunov exponents, \textit{i.e.}, the cumulative divergence rate of the trajectories, then makes clear the link between chaos and information theory.

In contrast with conservative systems, where Liouville's theorem applies, for dissipative systems such as fluid turbulence, phase space volumes are not constant along trajectories and are instead found to contract as they move along the trajectory in phase space. However, due to the presence of a subset of positive Lyapunov exponents, there can exist sub-volumes, \textit{i.e.}, spaces of lower dimensionality than the overall state space, which do preserve their volume along trajectories. This leads to the idea of an attractor, which is a subset of the state space to which the system is said to evolve towards. The dimension of this attractor can also be related to the Lyapunov spectrum via the Kaplan-Yorke conjecture \cite{kaplan1979functional,frederickson1983liapunov}, as
\begin{equation}
    \text{dim}(A) = j + \frac{\sum_{i = 1}^j \lambda_i}{|\lambda_{j+1}|} \ , \nonumber
\end{equation}
where $j$ is determined by
\begin{equation}
    \sum_{i = 1}^j \lambda_i > 0 \ \text{and} \ \sum_{i = 1}^{j + 1} \lambda_i < 0 \ . \nonumber
\end{equation} For chaotic systems $\text{dim}(A)$ is expected to be non-integer and the resulting attractor is referred to as a strange attractor.

\section{Theoretical predictions}

Multiple theoretical predictions have been made about the Lyapunov exponents and other chaotic measures in turbulence and specifically how they relate to other measures of turbulence.
We review some of these below.

\subsection{Theory of Lorenz}

Since some of the earliest observations of chaos were found in atmospheric 
models by Lorenz \cite{lorenz1963deterministic}, 
and atmospheric dynamics are strongly influenced by turbulence,
it is only fitting that predictions about the level of predictability of turbulence
immediately arose from these papers.

Lorenz considered a fluid with energy spectra
\begin{equation}
E(k) = A k^{-n} \ . \nonumber
\end{equation}
At each scale $k$, he posited a local time scale $\tau(k)$, which is proportional to the
turnover time of those eddies, which have wave length $2\pi/k$.
On dimensional grounds, this time scale is \cite{lorenz1969predictability}
\begin{equation}
\tau(k) = A^{-1/2} k^{(n-3)/2} \ , \nonumber
\end{equation}
and Lorenz then suggests that this is the time it will take for a perturbation to
grow from perturbation scale $k$ to $2k$.
In sum, this implies \cite{bohr1998dynamical}
\begin{equation}
\tau(k) \sim \frac{1}{kv_k} , \ v_k^2 \sim \int_k^{2k} E(k) dk \ , \nonumber
\end{equation}
with $v_k$ being the typical velocity difference. 
If we consider that $\tau(k)$ is minimised when $kv_k$ is maximised, 
which is also when $k^2 v_k^2$ is maximised, the $\tau(k)$ should be
dominated by the peak of $E(k)k^3$. 
This is different to the theory proposed by Ruelle below that the error is 
dominated by the Kolmogorov scale eddies.

Because this was derived within the context of atmospheric prediction,
it was immediately recognised that the degrees of freedom in the system could propagate errors in a way that was more important than the divergence of close trajectories.
Specifically, there could be different timescales that would grow the error at different rates and interact non-linearly. Because of these, it was expected that the predictability of the entire system did not have the maximum Lyapunov exponent as the key determiner of predictability. Indeed, it was found that the largest length scales had predictability times independent of the Reynolds number \cite{bohr1998dynamical}.




\subsection{Theory of Ruelle}

An alternative approach is that of Ruelle \cite{ruelle1979microscopic}.
He defines a characteristic exponent for the exponential growth of a disturbance $\Delta$
\begin{equation}
\lambda = \lim_{t \rightarrow \infty} \frac{1}{t} \frac{\ln \Delta(t)}{\ln \Delta (0)} \ . \nonumber
\end{equation}
He then reasons that this should depend on eddy size $l$. Using dimensional considerations
he derives
\begin{equation}
\lambda(l) \sim \varepsilon^{1/3} l^{-2/3} \ , \nonumber
\end{equation}
where $\varepsilon$ is the dissipation rate. 
He then states that the largest characteristic exponent should be associated with 
the smallest eddies, which are the Kolmogorov scale eddies 
$\eta \sim (\nu^3/\varepsilon)^{1/4}$, where $\nu$ is the viscosity. 
Overall, this implies
\begin{equation}
\lambda(l) \sim \frac{\varepsilon^{1/3}}{\eta^{2/3}} \sim \frac{\varepsilon^{1/2}}{\nu^{1/2}}
\sim \frac{1}{\tau} \ , \nonumber
\end{equation}
which in essence implies that the chaos is dominated by the smallest time scale, which
is the Kolmogorov time scale.
This theory is fundamentally based on the Eulerian sense of chaos, but has often been used in a Lagrangian sense. Comment is made upon this discrepancy in \cite{fouxon2021reynolds}.

\subsection{Multifractal corrections}

The above Ruelle theory implicitly relies on the K41 theory.
An attempted correction is made in \cite{crisanti1993intermittency} to account for intermittency effects.
Using the Kolmogorov theory, we have the prediction that
\begin{equation}
\eta \sim L \text{Re}^{-3/4} \ , \nonumber
\end{equation}
where $L$ is the integral length scale, and this implies
\begin{equation}
\lambda \sim \frac{1}{T_0} \text{Re}^{1/2} \ . \nonumber
\end{equation}
However, the turnover time of an eddy of size $l$ should be
\begin{equation}
\tau(l) \simeq \frac{v(l)}{l} \simeq T_0 l^{1-h} \ , \nonumber
\end{equation}
where $v(l)$ is the velocity of the eddy and
$h$ is the H\"older exponent of the velocity difference
\begin{equation}
\delta v(l) = |v(x + r) - v(x)| \sim V l^h \ , \nonumber
\end{equation}
and $l = r/L$ is some scaling parameter.
This implies that 
\begin{equation}
\eta \sim L \text{Re}^{-1/(1+h)} \ , \nonumber
\end{equation}
and so, 
\begin{equation}
 \lambda \sim \frac{1}{\tau} \sim \frac{1}{T_0} Re^{\alpha} \ , 
\ \alpha = \frac{1-h}{1+h} \ . \nonumber
\label{eq:twoalpha}
\end{equation}
The prediction $\alpha = 1/2$ is then a consequence of the Kolmogorov theory.
If $\zeta_n$ is the scaling exponent for the structure function $S_n(r) \sim r^{\zeta_n}$,
then in the Kolmogorov theory, 
$\zeta_n = n/3$ hence $h = 1/3$ and so $\alpha$ is predicted to be $1/2$. Differences from this value can then come from intermittency effects, where it is shown with a shell model of turbulence that $\alpha < 1/2$.
However, in \cite{fouxon2021reynolds}, Fouxon et.\ al.\ state that this derivation is wrong, since it relies on a questionable averaging procedure. In summary, the derivation relies on the regions with reduced chaos becoming larger due to intermittency, but the chaos is actually dominated by the remaining regions which have increased chaos, and contribute more to $\lambda$. Such structural behaviour is found in \cite{mohan2017scaling}.

\subsection{Predictions for attractor dimension and KS entropy}

Within the literature there exist a number of complex and mathematically sophisticated predictions for the scaling behavior of the KS-entropy and the attractor dimension for isotropic turbulence. However, simplified expressions can be obtained by considering the K41 theory and thus neglecting potential intermittency corrections. 

Considering first the attractor dimension, a dimensional argument was given by Landau and Lifshitz \cite{landau1959lifshitz} which estimated the degrees of freedom of a turbulent flow as the ratio of the largest to the smallest length scales. This is expressed as \begin{equation} 
\text{dim}(A) \sim \left(\frac{L}{\eta}\right)^3 \sim Re^{9/4} \ .  \nonumber
\end{equation} In \cite{ruelle1982large} Ruelle derived a more general result which allowed for a spatially intermittent dissipation rate, however, this requires detailed knowledge of the spatial probability distribution for the dissipation rate which is to date unknown. Instead, if K41 is assumed this result is consistent with the simple dimensional arguments.   Further estimates of scaling of the attractor dimension  can be found in \cite{constantin1985determining,gibbon1997attractor}. Finally, the multi-fractal model can also be used to derive a scaling prediction with intermittency corrections \cite{meneveau1989attractor}. Here the scaling is still in terms of the Reynolds number, however, the exponent is lowered compared with the K41 case.

For the KS entropy, fewer predictions have been made for isotropic turbulence. Ruelle \cite{ruelle1982large} again provided a result that can account for spatial intermittency, provided statistical information about the dissipation rate. Once more, if K41 is assumed then this result can be simplified and predicts \begin{equation} 
h_{KS} \sim \frac{1}{\tau_{\eta}} \bigg( \frac{L}{\eta} \bigg)^3 \sim  \frac{1}{T} Re^{11/4} \ .  \nonumber
\end{equation} Interestingly, this result is simply the product of the scaling predictions for the maximum Lyapunov exponent and the attractor dimension.



\subsection{Two Dimensional Turbulence}

In the above sections, we have focused on theoretical predictions for three-dimensional turbulence, however, the same ideas can also be applied to the two-dimensional case. The inverse cascade of energy in two-dimensional turbulence \cite{batchelor1969computation,kraichnan1967inertial,leith1968diffusion} has been found to lead to the formation of long lived and predictable coherent structures which has profound effects for predictability. 

For the maximum Lyapunov exponent, just as in three-dimensions, we follow the dimensional argument put forward by Ruelle that suggests it should be inversely proportional to the fastest timescale of the flow. We begin by assuming that in the direct cascade of enstrophy the energy spectrum takes the form $E(k) \sim k^{-3}$ ignoring, at this stage, any logarithmic corrections. Importantly, this form suggests that there is only a single timescale, $\tau$ associated with the entire forward cascade which is set by the enstrophy dissipation rate, $\eta$, such that $\tau \sim \eta^{-1/3}$. This then implies \begin{equation} 
    \lambda_1 \sim \frac{1}{\tau} \sim \eta^{1/3}.  \nonumber
\end{equation} The enstrophy dissipation rate is controlled by the enstrophy injection scale at the top of the cascade. Interestingly, in contrast with three-dimensional turbulence this suggests that in two-dimensional turbulence the smallest scales of the flow contain information about the large length scales that generated them.

Applying similar dimensional arguments to the KS entropy we can also estimate its scaling behavior. In line with the three-dimensional case, we expect the entropy scaling to be given as the product of the inverse of the fastest timescale and the total number of excited modes, the former we know from the maximal exponent and the latter will give the scaling of the attractor dimension. We consider the ratio of the largest and smallest scales of the flow and use this as an estimate of the scaling of the attractor dimension. In doing so we require the dissipation length-scale, $\chi = (\frac{\nu^3}{\eta})^{1/6}$ where $\nu$ is the viscosity. The ratio then implies \begin{equation} 
    \text{dim}(A) \sim \bigg( \frac{L}{\chi} \bigg)^2 \sim \text{Re},  \nonumber
\end{equation} and hence the KS entropy should scale as \begin{equation} 
    h_{KS} \sim \frac{1}{\tau} \text{Re} = \eta^{1/3} ~\text{Re}. \nonumber
\end{equation}

The above picture is incomplete since it ignores the logarithmic corrections to the energy spectrum that were suggested by Kraichnan \cite{kraichnan1971inertial}. These corrections are necessary for a constant enstrophy flux in the direct cascade. The influence of this logarithmic correction on the chaotic properties of two-dimensional turbulence were investigated by Ohkitani \cite{ohkitani1989log} and modify the above scaling predictions \begin{equation}
    \lambda \sim (\eta \text{log(Re)})^{1/3}, \nonumber
\end{equation}
\begin{equation}
    \text{dim}(A) \sim \text{Re} (\text{log(Re)})^{1/3}, \nonumber
\end{equation}
and
\begin{equation}
    h_{KS} \sim \eta^{1/3}\text{Re} (\text{log(Re)})^{2/3}. \nonumber
\end{equation} 

Ruelle's predictions in \cite{ruelle1982large} were modified for the two-dimensional case in \cite{lieb1984characteristic} by Lieb.
Whilst it is suggested that during the inverse cascade there is intermittency in the energy dissipation rate, in the absence of this effect, the scaling for entropy is \begin{equation}
    h_{KS} \sim \frac{\varepsilon}{\nu^2} V, \nonumber
\end{equation}
where $V$ is the volume of the system, whereas for the attractor dimension it was found that \begin{equation} 
    \text{dim}(A) \sim \sqrt{\frac{\varepsilon}{\nu^3}} V. \nonumber
\end{equation} These are consistent with the picture suggested by the dimensional argument in which the chaotic properties are influenced by all scales in the system.

\section{Results from simulations}

The above theory has been tested using numerical simulations.
In discussing the results of simulations, we have split the discussion into various categories. However, they interlink and it is impossible to create a totally clear division between them.
The main ideas that come up again and again are; how the chaos depends on the turbulent statistics, the different length scales that exist in the way that chaos presents in turbulence, and how this can allow more efficient simulation.

\subsection{Assessing Theoretical Predictions}

\subsubsection{Ruelle's prediction}

\begin{figure}
\includegraphics[width=0.49\textwidth]{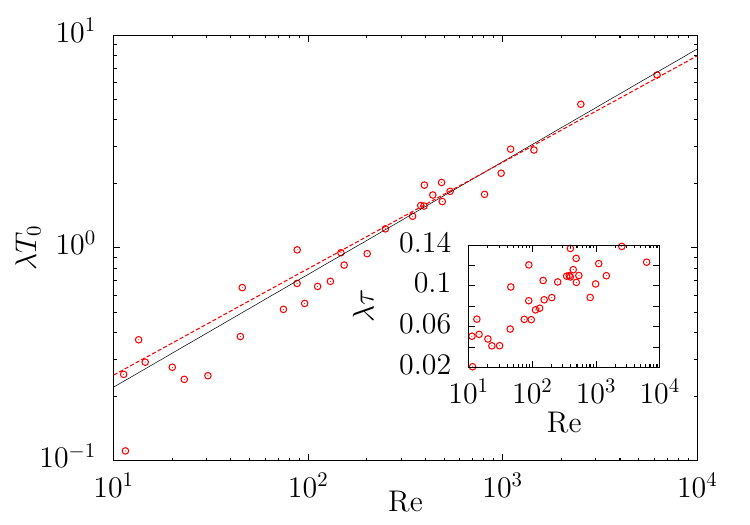}
\caption{Re against $\lambda T_0$ with dashed red line showing fit of $\alpha = 0.53$ whilst solid black line shows $\alpha = 0.5$. Inset shows a direct test of constant $\lambda \tau$. Figure taken from \cite{berera2018chaotic}.
\label{fig1}}
\end{figure}   

Multiple simulations into the Eulerian characteristics of chaos in turbulence, including those by us, have shown that Ruelle's prediction of constant $\lambda \tau$ does not hold, and instead that it increases with Re \cite{berera2018chaotic,mohan2017scaling,boffetta2017chaos}. These simulations reached as high as $2048^3$ collocation points \cite{berera2018chaotic} by us and Re = 8224 \cite{boffetta2017chaos} by Boffetta \& Musacchio. These three papers had slightly different emphases.
In \cite{mohan2017scaling}, Mohan et.\ al.\ explicitly measured the dependence of $\lambda \tau$ on Re$_\lambda$, with a dependence $\lambda \tau \sim$ Re$_\lambda^\beta$. This exponent was estimated with Baesian inference finding that the most likely value had
$1/4 < \beta < 1/3$, "with the possibility that the value is zero essentially precluded."
In our work \cite{berera2018chaotic}, the relation of $\lambda T_0$ to Re was measured, where we found $\lambda T_0 \sim C $Re$^\alpha$, and $\alpha = 0.53 \pm 0.03$. 
This fit was done to directly compare it to previous simulations involving shell models and predictions of intermittency using the H\"older exponent, which predicted that the value for $\alpha$ would go below 1/2 \cite{aurell1996growth}.
In Boffetta \& Musacchio \cite{boffetta2017chaos}, the relation of $\lambda T_{E0}$ to Re was measured, where $T_{E0} = E/\varepsilon$, where it was found that $\lambda T_{E0} \sim$ Re$^\beta$, with $\beta = 0.64 \pm 0.05$. They find that $\lambda \tau$ being constant is a poor assumption.
This exponent was significantly different to that in \cite{berera2018chaotic}, but this discrepancy was resolved in \cite{ho2020fluctuations}, which concluded that the difference between the value of 0.64 and 0.53 could be accounted for by the different timescales used, namely $E/\varepsilon$ for $\beta$ and $L/U$ for $\alpha$.

Interestingly, an even earlier paper by Mukherjee et.\ al.\ looking at predictability in dry convective boundary layers using large eddy simulation (LES), found a scaling equivalent to the prediction of $\lambda \sim $Re$^{1/2}$ with an exponent possibly even greater than $1/2$ \cite{mukherjee2016predictability}. These simulations were achieved using a simplified LES subgrid, which is similar to DNS. However, the direct Ruelle relation of $\lambda \sim 1/\tau$ was not directly measured.
These simulations also showed growth rate independent of perturbation length scale, and a self similar spectral shape, also discussed in depth below.

Whilst the simulations looking directly at the Ruelle prediction all showed non-constant $\lambda \tau$, in \cite{budanur2022scale} Budanur \& Kantz used simulations with very high resolution (meaning that $k_{max} \eta$ was much much larger than unity), and stated that the reason for an increase of $\lambda \tau$ with Re could be due to the unresolved scales in the previous simulations, since they found no increase of $\lambda \tau$ with Re. It was also stated that the artificial forcing used in these simulations could have resulted in non-physical Lyapunov scaling.
However, explicit tests by Mukherjee et.\ al.\ using identical Reynolds numbers found no dependence on resolution in error growth rates \cite{mukherjee2016predictability}.

Since the inverse of the maximum Lyapunov exponent can be associated with a timescale, a continuing increase of $\lambda \tau$ with Re will eventually lead to $\lambda \tau > 1$, which implies that there will be timescales in the system faster than the Kolmogorov microscale time.
Whilst it would be interesting to discover what happens at such scales in DNS, the observed very slow scaling of $\lambda \tau$ with Re means that this would require Re on the order of $10^9$, which is outside the realm of possibility for simulation in the foreseeable future.
If such a scaling did hold, it might imply that the instabilities act at even smaller scales than the Kolmogorov microscale, which is below the scale thought to be the smallest relevant scale in turbulence. Sub-Kolmogorov scales would then be of importance, and this should be tested as suggested in \cite{budanur2022scale} by Budanur \& Kantz. 
Although we believe the simulations performed in \cite{berera2018chaotic,mohan2017scaling,boffetta2017chaos}
are robust, this does not preclude the fact that new and interesting results can be found by testing this relation further.
Indeed, it has been suggested that thermal noises, which happen at smaller scales than the Kolmogorov microscale, may strongly affect predictability \cite{bandak2024spontaneous}.
Interestingly, in thermalised fluids, a link between the maximum Lyapunov exponents and the temperature has already been seen in many body chaos, suggesting an underlying universality \cite{murugan2021many}.

When observing the fit to data, the fit of $\lambda T_{E0}$ to Re$^\alpha$ is excellent \cite{boffetta2017chaos,ho2020fluctuations}, with scaling that does not vary greatly over the entire parameter range.
This begs the question, perhaps this relation to Re is the real one?
If it were, this would imply that $\lambda \tau \sim Re^{\alpha - 1/2} \sqrt{C(Re)}$, where $C(Re) = \varepsilon L/U^3$ is the dimensionless dissipation rate. In principle at low Re this may depend on Re \cite{mccomb2015energy}.
From the numbers found in \cite{boffetta2017chaos},
and since Re$_\lambda \sim$ Re$^{1/2}$, this would imply that $\lambda \tau \sim$ Re$_\lambda^{0.28}$, which is within the range calculated in \cite{mohan2017scaling}, despite these being independently measured.
The argument that $\tau$ is the relevant timescale of chaos is based on a simple physical argument, namely that chaos should be dominated by the smallest timescale of the system \cite{ruelle1979microscopic}, although care needs to be taken in this interpretation, as noted by Fouxon et.\ al.\ in \cite{fouxon2021reynolds}.
But, we know that chaos is a multiscale phenomenon, and it stands to reason that chaos in turbulence may also be a multiscale phenomenon. This is supported by the fact that the predictability of the large length scales acts fundamentally differently to that of the smallest scales. As such, it could be thought unlikely that the level of chaos in turbulence depends only on a small length scale quantity.

If the chaos is not totally derived from the smallest turbulence scale, the question is raised about where it does come from.
A possible resolution to this question is provided by Mohan et.\ al.\ in \cite{mohan2017scaling} itself, where it is discussed what the spatial contributions to the error were. The discussion there is particularly enlightening and should be read in full by the interested reader.
Here it was found, by looking at times where the field displayed extreme values of FTLE, that the difference between the two fields, ie where the error was located spatially, was dominated by local disturbances that were long lived.
In effect, the error growth globally can be dominated by large coherent structures in the error itself.
The average growth will be dominated by these extreme values and so they are particular important.
Within this region of high error, "the base field exhibits a pair
of corotating vortex tubes" \cite{mohan2017scaling}, which suggests that the growth of error is fundamentally linked to vortices in the flow, and would thus be affected by the rate of vortex stretching. This idea is further explored later in the review.
This effect of error growth being dominated by the vortices has also been seen in two-dimensional decaying turbulence \cite{boffetta1997predictability,ge2023production}.

The interpretation of $\lambda \tau$ increasing with Re is that it might be more realistic to expect that, as the flow becomes more extreme, there are more extreme regions that affect the maximum Lyapunov exponent, and since the extremes are non-Gaussian, and this extremity increases with Re, then the value should go up with Reynolds number.
This does not require there to be timescales that act on average faster than the Kolmogorov microscale time.

Because of the impossibility of directly testing Re where we expect that $\lambda \tau > 1$, we need to turn to theory, which can also provide exciting new directions for numerical studies of turbulence.
An example of such an intriguing theoretical contribution is found in \cite{fouxon2021reynolds}.
Fouxon et.\ al.\ note that the Ruelle relation is reliant on a lack of intermittency, whereas the Kolmogorov microscale time can change dependent on location.
Applied to fluid particles, this implies that there is a dependence on the location for fluid particles and so the two relations will not necessarily be the same.
Furthermore, the Ruelle relation was derived for turbulence, but has seen successful application to fluid particles embedded in turbulence. They show how one prediction can apply to the other.
The statistics of a field and the particles in that field are not necessarily the same, and are dependent on how statistics are calculated, and thus there needs to be a careful consideration of these which can prove fruitful theoretically and in simulations. This would have implications on any applicability of $\lambda \tau \sim$ Re$^\alpha$. These are promising avenues to pursue.
The authors also clarify the nature of intermittency in turbulence, stating that there are in fact at least two sources.
The first is due to random fluctuations in the field that exist even with Gaussian fields. The second is due to the intermittent nature of turbulence, which they define as "atypically large persistent velocity gradients" \cite{fouxon2021reynolds}.

\subsubsection{KS Entropy and the Lyapunov Spectrum}

\begin{figure}
\includegraphics[width=0.49\textwidth]{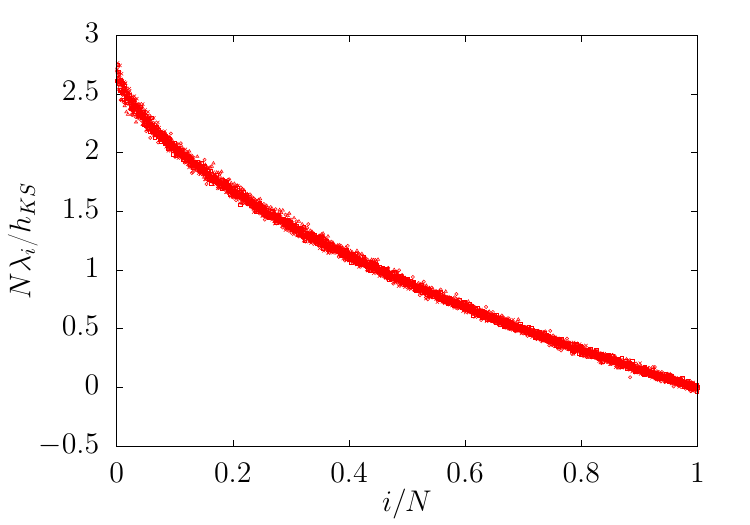}
\caption{Lyapunov spectrum for Re from 90 to 212, rescaled by $h_{KS}$ and number of positive exponents. Scaling is indicated by the collapse onto the same curve. Figure taken from \cite{berera2019information}. \label{fig2}}
\end{figure}

Extending beyond the maximum Lyapunov exponent, a robust use of chaotic
measures would require computing the spectrum of Lyapunov exponents,
which would be an alternative complete description of turbulence
in distinction to spectral measures.  This is a numerically intensive
problem.  The first attempt at such simulations were done
by us \cite{berera2019information} and independently in \cite{hassanaly2019lyapunov,hassanaly2019numerical} by Hassanaly \& Raman.

In \cite{berera2019information} we directly computed the Lyapunov spectrum for three-dimensional HIT using DNS. This allowed numerically rigorous tests of the mathematical relations and conjectures in the literature.
Since a very large spectrum was obtained, it allowed calculation of the KS entropy and thus the rate of information production.
The dimension is shown to be very large, even at low Re, being as high as 5704 for Re = 212.
Thus, low dimensional approximations will give poor estimates.
The results allow a direct comparison to the results of Ruelle, intermittency, and finite Reynolds number effects, given the model independent results.
The calculations are achieved using the Benettin algorithm, which is very costly, involving M FTLEs, one for each calculated exponent, and a Gram-Schmidt algorithm to orthogonalise the measurements such that a well ordered spectrum can be obtained, which meant that Re needed to be modest, ranging from Re 50 to 212, and M has to scale very quickly to capture all positive exponents.
We found that $h_{KS} T = C$Re$^\alpha$ with $C = 0.0008 \pm 0.0002$ and $\alpha = 2.65 \pm 0.06$, with commentary on agreement with predictions and problems with using shell models. However, the low values of Re tested place a large uncertainty in the scaling exponent. The shape of the spectrum was then compared to theory, and here a divergence density of exponents around $\lambda \approx 0$ that had been predicted by Ruelle \cite{ruelle1982large} was not observed. An Re independent shape for the spectra was found consistent with what was seen in  Poiseuille flow by Keefe et.\ al.\ \cite{keefe1992dimension}.

Finally, an estimate of the attractor dimension for a given Re was made. For the lowest Re cases this was calculated directly using the Kaplan-Yorke conjecture whereas for the higher Re cases this relied on the shape similarity property of the spectrum to enable a reliable prediction of the crossing point for the sum of exponents. The dimension is extremely high.
Landau and Lifshitz give a good approximation, with others compared for the number of positive exponents, finding $N = b$Re$^\gamma$ with $b = 0.008 \pm 0.002$ and $\gamma = 2.35 \pm 0.05$. 


We performed a similar set of simulations for two-dimensional turbulence in \cite{clark2020chaos},
which extended the results in a systematic way, pointing out its greater applicability to thin layer turbulence such as in the atmosphere.
There are different timescales in two- and three-dimensions, which results in different Re scaling. The simulations use large length scale dissipation to prevent the formation of a condensate at low $k$, and a stochastic forcing.
We found very good agreement for $\lambda_1 = 0.42 \eta^{1/3}$, which agrees with prediction, although the fit was done by varying the prefactor since varying the exponent in this case would not be dimensionally consistent. We then show that some correction to $\lambda \tau$ is required, resulting in some Re dependence, as suggested by Kraichnan.

In two-dimensions, it was found that in addition there is an influence on the forcing length scale and system size, suggesting non-universal behaviour in addition to what had been predicted from the Kraichnan theory.
The attractor dimension grew with the inverse cascade inertial range, but had a contribution from enstrophy.
As in three-dimensions, no divergence around $\lambda = 0$ in the spectrum is seen.
It was further found that dimensional arguments were not sufficient to account for predictions, with corrections based on forcing and length scale, due to the non-local effects of coherent vorticies. This again shows how important vortices are for the error growth.

A further study was performed in \cite{hassanaly2019lyapunov}.
Hassanaly \& Raman note how a dynamical systems understanding of turbulence can act as a complement to the statistical view of averages and deviations from the average. Whilst there are statistical quantities to capture these effects, their influence on the dynamics of the system can be illuminated with a dynamical systems approach.
In the study, they calculate the Lyapunov spectrum for three-dimensional HIT using low mach number incompressible flows with constant density. They calculate the Kaplan-Yorke dimension as $D_{KY} = (L/\eta)^{2.8 \pm 0.095}$.
From their plots of the Lyapunov vectors, the authors conclude that the vectors "themselves are disorganized fields, containing structures similar to the original velocity field".
After performing a statistical analysis between the velocity field and the LV field they find results similar to \cite{mohan2017scaling}, where the contribution to the error growth is contained in very small areas, furthermore finding that this fraction decreases with increasing Re. Hassanaly \& Raman found "a strong correlation between the chaotic parts of the domain and the velocity gradients at low Reynolds number. At higher Reynolds numbers, enstrophy becomes a good marker of chaoticity."
Lyapunov exponent and Lyapunov vector analysis has also been applied in \cite{hassanaly2019numerical} by Hassanaly \& Raman, and to magnetohydrodynamic (MHD) turbulence on a Shannon information entropy level  by Vasey et.\ al.\ in \cite{vasey2023influence}.

The full spectrum of Lyapunov exponents would be an alternative
complete description of the fluid flow.
Our studies showing the stability of such quantities for the maximal
exponent, we expect also holds for the other exponents.
Thus computing these exponents may prove to have considerable
numerical advantages. We did an initial examination of
such a calculation in \cite{berera2019information} and it seems beneficial to develop such studies in the future.

\subsection{Predictability}

\subsubsection{Error spectrum}

\begin{figure}
\includegraphics[width=0.49\textwidth]{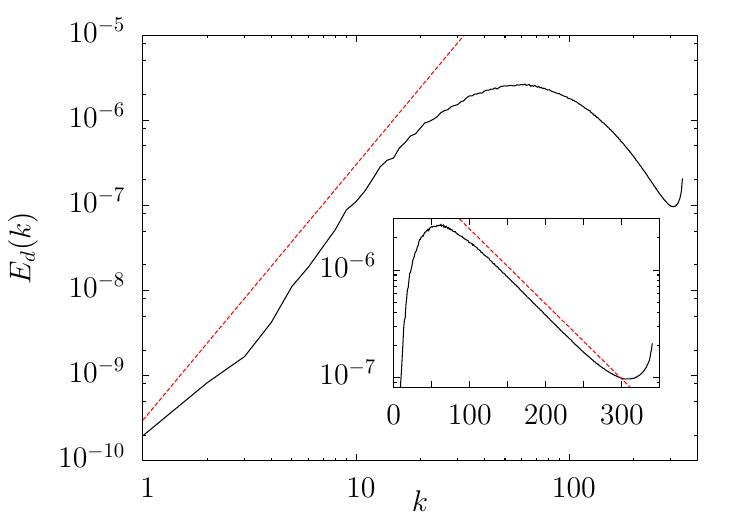}
\caption{Characteristic error spectrum $E_d(k)$, shown for Re = 2500 with log-log main figure, and semi-log inset. Main figure has dashed line $k^3$, inset has exponential.  Figure taken from \cite{berera2018chaotic}. \label{fig3}}
\end{figure}

By looking directly at the error spectrum we can see the spatial distribution of the error, in a similar way to looking at it visually.
By following the direct approach, we get to see the different scale behaviour, as opposed to FTLE \cite{berera2018chaotic}. It is harder to see the different scale behaviour in FTLEs since those saturate at exponential growth regimes, the spectral information is lost, and since different scales are important for different problems. The "naive" approach has validity, especially since we can directly probe different scales (see the eddy damped quasi-normal Markovian (EDQNM) approximation for an important example).
We looked at the difference spectrum in DNS in \cite{berera2018chaotic}. There is a power law dependence in the small wavenunmber region where the exponent grows as 3 or 4, and then drops off exponentially beyond the peak, which occurs at roughly $k_\eta$, the wavenumber of the Taylor microscale. As Re increases, this exponent becomes increasingly flat (a white spectrum) and indicates that, below that scale, the noise is distributed at all scales equally.
Further exploration shows that, using the error spectrum, we see dynamical changes that depend on the saturation of the error spectrum.
The transition from exponential to linear growth happens when the error saturation occurs at $k_\eta$ \cite{boffetta2017chaos}, with the FSLEs also giving important structural information, finding "the time for the perturbation to affect a wave number k in the inertial range is proportional to $\varepsilon^{-1/3} k^{-2/3}$."

In \cite{yoshimatsu2019error}, Yoshimatsu \& Ariki analyse an initially high-k error in three-dimensional turbulence and find similar results. 
They use self-similarity arguments to find an integral length scale of the error field and show how the normalised error spectra collapse, and in the low wavenumber range find a $k^4$ dependence.
They use a spectral cutoff filter to split the error production into high and low wavenumbers dependent on a critical wavenumber, finding that the interaction of the two contributes to the error scaling at low wavenumber, whereas the growth of the absolute value of the field is mainly provided by the low wavenumbers. This goes against the common idea that the error growth is driven mainly by the smallest scale features which propagate upwards. Instead, we have the large length scale features being important as well.
One of the advantages of DNS is being able to probe such realisations and understand turbulence better, and chaotic measures are vital for this understanding.

\subsubsection{Cascades of error}

In \cite{lorenz1969predictability}, Lorenz was primarily concerned with the way that an error in the small length scale
could propagate upwards in scale, so that a lack of knowledge at small length scales would imply limits on predictions of the large length scales.
However, it has been pointed out that, when applying the analysis to a turbulent fluid, there are drawbacks since the multiscale error interacts not just in an inverse cascade, but forwards as well.
This is partially analogous to the situation of energy cascades in turbulence.

In Fourier space, Waleffe analysed turbulence by its fundamental triadic interactions. Under the assumption of statistical homogeneity, the nonlinearity of the Navier-Stokes equations can be understood in Fourier space as the interactions of triads of wavevectors $k, p, q,$ such that $k + p + q = 0$. Formally, each such
triad of wavevectors defines a dynamical system describing the interaction of the corresponding Fourier modes, which makes it possible to investigate the behaviour of different interactions between helical Fourier modes.
Most attention was historically spent on analysing the forwards cascade of energy, however certain triadic interactions of turbulence actually sum to an inverse cascade of energy \cite{waleffe1992nature}, as has been shown in numerical simulation \cite{biferale2012inverse}.
Such triadic decompositions in MHD also display intriguing behaviour
\cite{linkmann2016helical}, which is relevant because MHD also displays novel chaotic behaviour compared to fluid turbulence as discussed below.
Similar insights could be gained by considering error through triadic interactions.
In \cite{rotunno2008generalization}, Rotunno \& Snyder generalised
the Lorenz model of propagation of errors upwards to "surface quasigeostrophic equations". Here it was found that the downscale error propagation is of vital importance for understanding the predictability dynamics,
and that the importance of the interaction of large and small length scale features for the error has also been noted as long ago as 1957 \cite{thompson1957uncertainty}.
It has also been shown that in weather forecasting, it is not only errors from the small scales that affect the predictability, but that large scale errors may be a limiting factor \cite{durran2014atmospheric}.

We looked at the cascade of error in \cite{berera2018chaotic} implicitly by testing perturbations made at different scales.
The perturbations were spatially isolated to largest scales for the main results, but tests were done with high wavenumber (large length scale) perturbations.
The simulations showed that that perturbations made at large length scales lead to an initial convergence of the two fields, suggesting that a portion of the error must cascade forwards to smaller scales in order to grow them, and that also there was a diminution of error. This was not the case for the perturbations made at small length scales (low wavenumbers). 
As well, exponential growth in the low wave number region was seen if forcing was performed at intermediate wavenumbers, suggesting
that the cascade of error is not necessarily dependent on the forwards cascade of energy, even if it does go forwards.
Further insights on the scale of error growth were made in \cite{clark2022critical}
and examined more explicitly by Ge et.\ al.\ in \cite{ge2023production}.

The fact that there is scale dependent error growth has important implications for predictability and simulations of LES \cite{budanur2022scale}. The scale dependence was one of the original motivators of the FSLE. In LES, if there is a scale dependent error growth, the prediction horizon will depend on the grid spacing. Numerical simulation on LES shows that as the resolution is increased, there is a trend towards the DNS value.
LES simulations show that the smaller the resolution, the faster that the perturbation/error can grow \cite{budanur2022scale}.
This tradeoff between predictability and error growth allows one to determine when "the point of diminishing returns is reached" and Budanur \& Kantz find that "the forecast horizon cannot be extended by simply improving resolution alone."

\subsubsection{Linear growth of error}

After the initial exponential growth phase, linear growth of error was seen in multiple simulations 
including ours \cite{berera2018chaotic,boffetta2017chaos}.
It was found to depend linearly on dissipation, and this was linked to the KS entropy. However, a possibility was mentioned that this could be forcing dependent \cite{boffetta2017chaos,ge2023production}.
At late stages, the predictability is dominated by these late stages, which allows for the long time predictability in atmospheric systems.
In MHD, as mentioned below, we further found  that the linear growth rate depended on the dissipation rate of the particular field \cite{ho2019chaotic}. Since in cosmological scale magnetic fields, the magnetic dissipation can be extremely low, this allows in principle extremely long prediction horizons even in turbulent electrically conducting fluids.

\subsubsection{FTLE statistics}

The FTLE are themselves a statistical quantity and can be analysed in a similar way to fluctuations of the energy and dissipation,
and looking at the instantaneous values of the FTLE reveals important details for the flows at that point in time \cite{mohan2017scaling}.
FTLE have been used to characterise predictability of weather events \cite{pandey2021short} and extreme values in geophysics models \cite{sterk2012predictability}.

A more systematic look at the FTLE was performed by us
in \cite{ho2020fluctuations}.
It was noted that the statistics of the FTLE are in some way better than using the kinetic energy and many other statistics since their autocorrelation time was much shorter than for any other analysed statistic or energy at any scale or dissipation, a finding similar to \cite{mohan2017scaling}.
The Eulerian maximum Lyapunov exponent is a robust measure of level of chaos, and gives better statistics than Reynolds number or energy, which means it may be a better characterisation of the system.
We believe this is because the deviation of trajectories is sampled at every grid point in the simulation, the number of which is always greater than the number of grid points contributing to any spectral quantity.
FTLEs are stable even for short steptimes for the FTLE,
as we found in this paper,
thus reducing computational cost needed to obtain large sets and improve statistics (the energy varies only as the large eddy turnover time, for instance). Whilst the direct method gives better knowledge of the spectrum, we receive less statistics for the maximum Lyapunov exponent itself and its fluctuation.
The simulations also showed that the FTLE gave steady statistics at a much shorter time after initialisation than Re, and E, which suggested that meaningful results may be gained from them much more quickly than these other quantities, which are more affected by hysteresis. Allowing less relaxation time for the simulation is also useful for efficient simulation, which as a topic is explored further below.
Contrary to the findings of Budanur \& Kantz \cite{budanur2022scale}, we found in these DNS simulations that the maximum Lyapunov exponent did not depend on the lattice size at all, even when the simulation was super well resolved.

Statistical analysis showed that the skewness of total energy, Re, and, dissipation (which is a small length scale quantity) are positive, whereas the skewness is negative for $\lambda$.
It was also found that the fluctuations of the maximum Lyapunov exponent were significant, and calculating them was not captured by a Gaussian propagation of errors. "$\sigma_\lambda / \lambda$ / fluctuations are significant and they cannot be simply estimated by measuring the values of Re, $\sigma_{Re}$, $T_0$, and $\sigma_{T_0}$" \cite{ho2020fluctuations}. This suggests that the extreme values of the FTLE, which can drive growth of error in the long term, cannot be determined simply.
The fluctuations of the Laypunov exponent increased with Re as a power law.
The rate of these fluctuations allows one to estimate the time needed to run a simulation for good statistics, and is thus an explicit way of attaining efficient simulation.
Specifically, they would allow us to predetermine over how many large eddy turnover times statistics need to be taken to be representative.
The fact that FTLEs are the most robust statistical measure we could find, gives impetus to the use of it as a standard measure in DNS of HIT, even working in MHD.

\subsection{Efficient simulation}

Chaos can be seen in turbulence even at very small box sizes, as was seen already using $32^3$ collocation points by Deissler in 1986 \cite{deissler1986navier}.
Whilst some chaotic properties may require heavy numerical work, using chaos as a measure can improve the efficiency of simulation, by directing resources where they would be most appropriately needed.
For instance, if only a certain tolerance is needed on predictability, it may be computationally wasteful to go over that tolerance by an order of magnitude. Understanding the chaotic measures of turbulence allows us to determine this point explicitly.


The fact that the predictability can be quantified was used to achieve more efficient simulations in \cite{wang2022synchronization}, where Wang \& Zaki used it to argue about the fact that the small length scale errors cannot contaminate the large length scales by the time it reaches downstream.
Whilst analysis of the Lyapunov spectrum in turbulence may seem of mainly theoretical interest, the application of the method to practical problems, such as analysing separated flow around an airfoil \cite{fernandez2017lyapunov}, shows that it can be a useful tool even in practical situations. In this study Fernandez \& Wang found a dependence on the mesh refinement, as seen elsewhere.

Large-eddy simulations work by a filtering of the Navier-Stokes equations \cite{leonard1975energy}, with a model for subgrid scales.
Within the context of LES, authors often use the term "error" to refer to the difference between a simulated value and the statistics that would have been obtained from a full simulation.
The maximum Lyapunov exponent has been used to characterise the growth of numerical error in LES simulations by Bae \& Lozano-Duran \cite{bae2022numerical}.

In \cite{nastac2017lyapunov} Nastac et.\ al.\ propose using Lyapunov exponents to evaluate LES quality by analysing the specific problem of a turbulent jet flame and turbulent combustion, and their effects on predictability.
They use the Eulerian field, finding that, whilst in the DNS regime there is no effect of grid size on the predictability, here there is, but only for inert flows.
When they looked at reaction flows, they found a different effect
and further studied how chaotic analysis can improve simulations as a metric for determining what are the relevant scales of a specific problem.
The Lyapunov exponent can determine regions of high error growth, and thus regions where grid resolution should be increased. The exponents can be used "to estimate the time required for upstream turbulence to propagate downstream."
The authors further note that for an inert jet, this is trivial, but can be useful for complex flows and geometries generating turbulent structures.
Chaotic measures analysis is a general method for LES and does not depend on the subgrid scale modeling. Once the limit of the maximum Lyapunov exponent is reached, it should approximate the chaotic dynamics accurately.
Naive application of the method showed the physics of the inert jet, namely that the most important region was close to the nozzle.
There was then more specific comment on how it applied to the physics of turbulent jets, and shows the strength of the application, especially to how combustion can make the dynamics more predictable, partly due to the flame relaminarisation.
This sort of analysis may be especially promising since measuring $\lambda$ is fairly simple. 

A different analysis of LES quality was performed in \cite{wu2018lyapunov}.
Wu et.\ al.\ pointed out how Lyapunov exponents are useful in transient simulations, where statistically stationary flow is not achieved. This is also useful for capturing rare events, typical in combustion, happening at short time scales.
The maximum Lyapunov exponent can be found for arbitrary geometries.
They find that once the dynamical scales are fully resolved, the maximum Lyapunov exponent saturates, and thus we know the maximum resolution needed for efficient simulation.
However, they warn that this cannot be known beforehand and requires repeated simulation to discover.
Even so, the authors conclude that these metrics are "capable of identifying quantity-specific sensitivities with respect to the numerical resolution, while requiring significantly less computational resources than acquiring profiles of conventional turbulent statistics."
This analysis was extended in Engelmann et.\ al.\ \cite{engelmann2022towards}, using the Shannon entropy in LES flows and its specific application to wall and shear flows. Here as well they find that Shannon entropy can be "a more stringent quality criterion than resolved energy" \cite{engelmann2022towards}.
In effect, information entropy can be a useful tool for assessing LES quality, not just statistics of energy.

This idea that we can use the entropy production (or other dynamical systems quantities) to analyse the suitability of simulations, can extend to other non-linear systems on unstructured grids (eg for biological simulations), and would provide a different metric for determining where computational resources should be spent efficiently. Of course, resolvability criteria are important, but this is another tool.



In engineering devices, one common problem encountered is to optimise some time averaged quantity. When done using least squares shadowing, this depends on the integration of a number of equations equal to the number of positive Lyapunov exponents. Thus, understanding the way that the Lyapunov spectrum changes with parameters, such as the Reynolds number, is of key importance. Various strategies have been proposed to surmount this problem \cite{kantarakias2023sensitivity}.
A recent theoretical framework of data assimilation approaches, depending on the chaotic properties of turbulence, allows one "to infer small-scale turbulent structures based on observational data from large-scale ones" \cite{inubushi2023characterizing}.
Use of Lyapunov analysis has also been done to provide predictions of combustion events in the context of sequential data assimilation \cite{magri2020physics}.

\subsection{Insights}

\subsubsection{Theoretical extensions}

\begin{figure}
\includegraphics[width=0.49\textwidth]{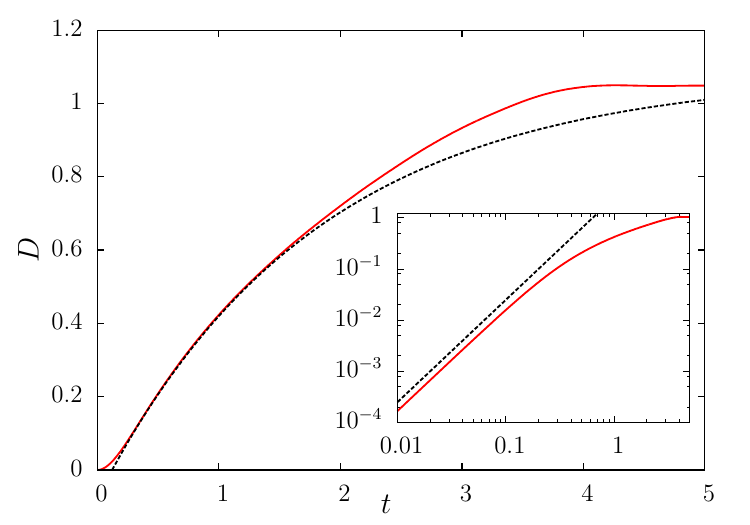}
\caption{Forgetting rate with time shown as change in drift energy $D$. The dashed line in main figure is the prediction $D = 2E(1 - \exp(-t/\tau)$ and in the inset it is $t^2$. Figure taken from \cite{ho2019high}. \label{fig4}}
\end{figure}

\begin{figure}
\includegraphics[width=0.49\textwidth]{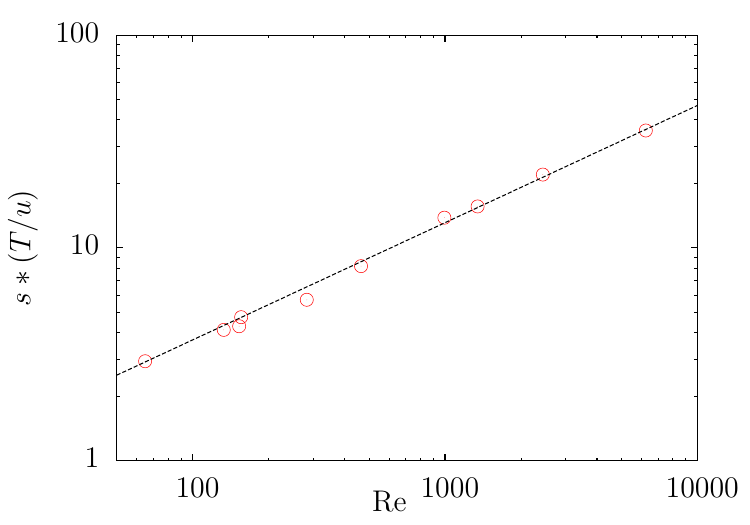}
\caption{Normalised phase space speed and Re. Interestingly this uses only large length scale quantities: $T$, the large eddy turnover time; and $u$, the rms velocity. Dashed line shows a least squares fit. \label{fig5}}
\end{figure}

The analysis of turbulence using DNS allows different mathematical theories to be tested numerically, such as the idea of superfast amplification and nonlinear saturation of perturbations \cite{li2020superfast}.
Lyapunov analysis of turbulence has also aided thoughts about stochasticity, such as in renomalisation group approaches by Eyink \& Bandak \cite{eyink2020renormalization}.
This paper understands turbulence from a more dynamical systems approach and the understanding of turbulence through the lens of chaos is intimately linked to this approach.
The question of how the "forgetting" is achieved can be understood by the drift over time, $D$, where
\begin{equation}
    D(t) = \frac{1}{2} \int (u(t) - u(t_0))^2 \ dV \ . \nonumber
\end{equation}
The behaviour for a sample run is shown in Figure~\ref{fig4} \cite{ho2019high}. As well, we can define the phase space speed, $s(t)$, where
\begin{equation}
    s(t)^2 = \frac{\int (u(t + \Delta_t) - u(t))^2 \ dV}{\Delta_t^2} \ ,   \nonumber
\end{equation}
with $\Delta_t$ being the simulation time step. The dependence on Re is shown in Figure~\ref{fig5}, which only depends on the large length scale quantities indicating that the majority of the speed in phase space is dominated by the large length scales. This can be expected by the fact that these scales contain the majority of the energy. 
That such easily obtainable quantities may be of use to theorists highlights that simulations of turbulence still contain rich veins to be drawn from, and that a quest for ever larger box sizes is not necessarily the best use of computational resources.

Further, analysis of the maximum Lyapunov exponent on active turbulence may reveal interesting things about predictability of active matter \cite{singh2022lagrangian}, which is important since these are key biological processes.
This analysis was performed in \cite{mukherjee2023intermittency},
where Mukherjee et.\ al.\ found different behaviour at different levels of activity.

\subsubsection{Changes in system dimension}

Changes in system dimension can change the chaotic properties of turbulence.
We were motivated \cite{berera2020homogeneous} by studies in critical phenomenon by Wilson \& Fisher
looking for a critical spatial dimension
above which the critical exponent is given by mean field
theory \cite{wilson1972critical}.
We wanted to discover if similar effects held in fluid turbulence.
It has been suggested
that K41 is in fact a mean field theory, exact only
above a critical dimension \cite{siggia1977origin,bell1978time}.
In our previous studies we found that chaotic properties
are highly dependent on the spatial dimensions
for two and three spatial dimensions \cite{berera2018chaotic,clark2020chaos,clark2021chaotic}.
We thus utilised such measures to
examine higher spatial dimensions using EDQNM and some DNS
\cite{clark2022critical}.

In \cite{clark2021chaotic}, we studied the transition from two-dimensional to three-dimensional turbulence using chaotic measures. These are very sharp measures as compared to looking at turbulence spectra, which are much more affected by noisy data. The study uses a box with decreasing height to study the transition, with a stochastic external forcing. As the box height decreases, the phenomenology of the system transitions from purely three-dimensional to an intermediate state of mixed two- and three-dimensional behavior and finally on to purely two-dimensional \cite{benavides2017critical, ecke20172d, van2019condensates, musacchio2017split}. The very different behavior of vorticity between two- and three-dimensions is likely part of the explanation for the sharpness of chaotic measures in this case. 

\begin{figure}
\includegraphics[width=0.49\textwidth]{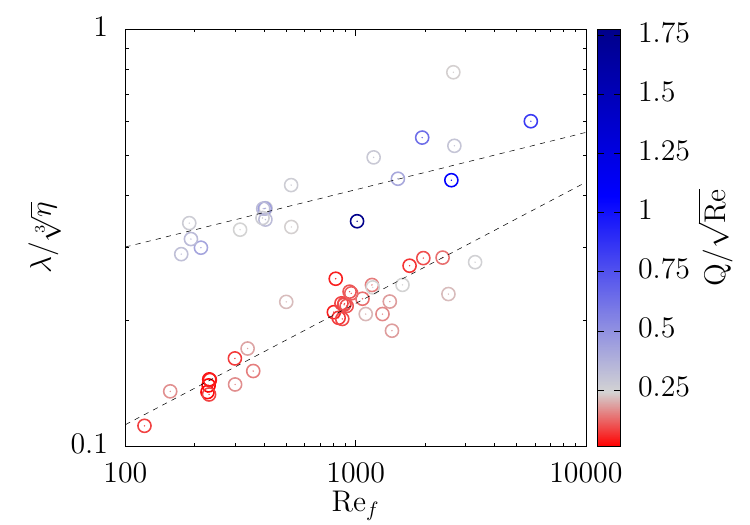}
\caption{Reynolds number scaling of the maximum Lyapunov exponent in thin-layer turbulence. A discontinuous jump is found as a function of the aspect ratio of the system. Figure taken from \cite{clark2021chaotic}. \label{figReTL}}
\end{figure}

\begin{figure}
\includegraphics[width=0.49\textwidth]{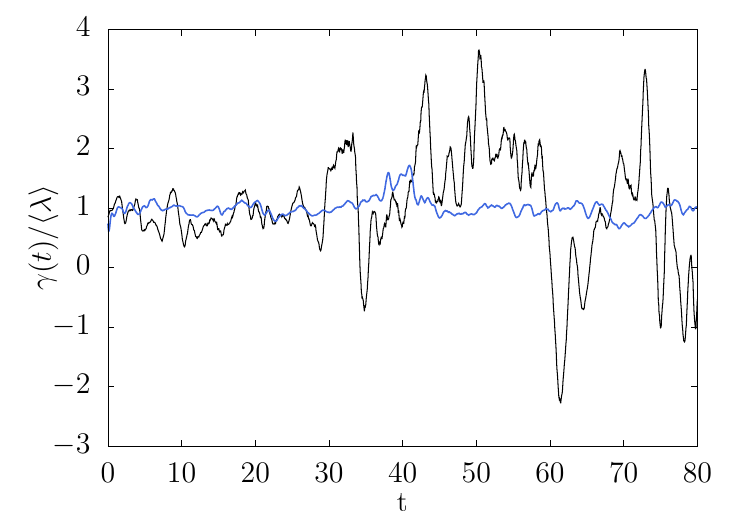}
\caption{Time history of the maximal FTLE computed near to the transition point (black) and far from the transition (blue). The statistics of the point near the transition displays large fluctuations. Figure taken from \cite{clark2021chaotic}. \label{figFLucTL}}
\end{figure}

We observed a discontinuous jump in $\lambda$ when properly scaled \cite{clark2021chaotic}. This was in contrast to more standard statistics which exhibited a smooth transition. Additionally the time history of the FTLE becomes more intermittent around the transition point showing large fluctuations, hence these chaotic measures really are capturing information about the flow. The observed discontinuity in the maximum Lyapunov exponent as a function of box height as well as the enhanced fluctuations around this point are reminiscent of a phase transition. This suggests chaotic measures can reveal additional information that is not accessible to standard statistical measurement. The use of chaos to study phase transitions has only seen limited discussion in the literature \cite{butera1987phase, caiani1997geometry, barre2001lyapunov}, however, our findings in \cite{clark2021chaotic} suggest this approach should be examined more carefully.
We also saw an interesting response to system dimension in EDQNM models \cite{clark2022critical}, which show strong dependence of the chaotic measures on dimension, discussed below.

\subsubsection{Importance of vortex stretching}

One of the themes that emerges from our and other studies into chaotic measures in turbulence is the importance of the vortical structures to the Eulerian chaos.
We emphasised this in \cite{clark2022critical}, which followed on from our previous work in \cite{clark2021effect} and looked specifically at the effect of spatial dimension on EQDNM models of turbulence.
The introduction of this paper contains a succinct overview of previous uses of EDQNM to study dimensionality effects and further references can be found therein.
Briefly, the EDQNM model starts by using the Navier-Stokes equations in Fourier space to derive an equation for the evolution of the third order velocity field moment.
A quasi-normal approximation is used to close the infinite moment hierarchy and further adapted, resulting in the eddy damped quasi-normal Markovian (EDQNM) approximation, as shown by Orszag \cite{orszag1970analytical}.
In this study, we then derived an equation for the
correlation spectrum between two different velocity fields under this closure which can then be used to measure the maximal Lyapunov exponent within the EDQNM closure.

The study was motivated by wanting to better understand criticality and scale invariance. Notably, EDQNM models provide an algebraic method by which to vary the system dimension resulting in little additional computational expense. Physically it was found that the enstrophy production peaked near six dimensions, resulting in changed dynamical behaviour.

Whilst the equations themselves are not chaotic, it is possible to deduce the error growth within the model \cite{orszag1970analytical,leith1972predictability}. 
We found that DNS and EDQNM give a similar order of results \cite{clark2022critical}, where DNS and EDQNM $\lambda$ decreases with dimension. The EDQNM runs showed that the Ruelle relation breaks down even in EDQNM in a way that is similar to DNS because of a lack of "scaling range and hence no separation between the viscous and large-eddy time scales".
Most interestingly, a critical dimension was found for error growth.
From analysis of the error spectrum, we can look at the predictability time and find good agreement with DNS, but we get some late time saturation even in 3D.
In terms of error cascades, there is a "competition between the generation of uncorrelated energy and its sweeping out by the cascade" \cite{clark2022critical}.
In the EDQNM runs, we see that this forward cascade saturates at a higher wavenumber. There is a cascade of correlated energy which is more effective with higher dimension.
It is speculated that the diminution of error is then caused by the change in the vortex stretching term as we increase dimensions.

In \cite{ge2023production}, Ge et.\ al.\ ran with these and other ideas to more explicitly analyse the effect of vortex stretching and compression on error growth, finding competition between them.
The authors derive the evolution equation for the error for HIT, finding "three different mechanisms: internal production
resulting from interactions between the strain rate and the velocity-difference field,
dissipation of the velocity-difference field and external force input."
Finding that, indeed, the production of error "can be related to the strain self-amplification and vortex-stretching".
They then clarify the effects that the forcing can have on the production of error.
They find that "spontaneous decorrelation of a flow from its perturbed flow in the absence of external inputs of uncertainty can only occur through local compressions",
and a self similar uncertainty spectrum similar to \cite{yoshimatsu2019error}.
This spectrum implies a characteristic length of uncertainty that is defined and tracked. The authors associate it as possibly being related to the Taylor length scale, and link this to the finding by Goto \& Vassilicos in \cite{goto2009dissipation}, where Ge et.\ al.\ state this to be "the mean distance between stagnation points in a HIT and therefore tends to represent the average size of turbulent eddies which is highly weighted towards the more numerous smallest ones."
They propose a large length scale random sweeping effect as being a possible driver of error growth, which is in agreement with the highly localised nature of the error found in \cite{mohan2017scaling}, and is motivated by their finding that the chaos is driven by events with extremely high kurtosis and positive skewness.
Finally, the authors conclude that "the identification of local compression and stretching events as key to the development of uncertainty means that future prediction methods may benefit from strategies for early detection of such events so as to concentrate maximum accuracy on the compression events and less accuracy on the stretching events." This again proves the necessity of the method for efficient simulation, and how different strands of the theory and results combine to mutually advance each other.

Whilst we emphasise chaotic measures as an alternative for analysing turbulent flow in this review, we do not believe that they are exclusive.
We have already seen how the error spectrum is a useful way of understanding the error, 
it would be useful to also analyse the error transfer, as derived in \cite{ge2023production}.
This can be spectrally decomposed and provide a measure of the transfer of error across scales in an analogous way to the transfer of energy in HIT.


\subsubsection{MHD}

We extended the study of Eulerian chaotic measures by tracking trajectories in phase space in turbulence to MHD in \cite{ho2019chaotic,ho2020fluctuations}, whilst a theoretical study by other Fouxon et.\ al.\ was conducted in \cite{fouxon2021reynolds}.
Shell models of turbulence applied to MHD with Prandtl number $\sim$ 1 found
$\lambda \sim \nu^{-1/2}$ \cite{grappin1986computation}, however, since shell models may not be appropriate for the study, we used DNS simulations.

In MHD a magnetic field is coupled to an electrically conducting fluid, with dynamics given by the MHD equations \cite{elsasser1950hydromagnetic,cramer1973magnetofluid}.
The equivalent of the Ruelle relation was complicated by the fact that there are multiple microscales in this system.
The non-linearity of the equation is direct for the velocity field, but for the magnetic field the non-linearity arises indirectly through a coupling with the velocity field.
Thus, we may expect that the chaos should be dominated by the velocity field and not the magnetic field, which was indeed found to be the case.
A test of removing the Lorentz force confirmed that it really is the velocity that drives the chaos \cite{ho2019chaotic}.
No stable trend was associated with the Prandtl number, though using $\lambda \sim 1/\tau$ was found to fit better due to the noise from the Prandtl number and the fact that the relationship between Re and $\tau$ is not as simple in MHD as for HD. For the relationship $\lambda T_0 \sim$ Re$^\alpha$ it was found $\alpha = 0.43 \pm 0.09$.

Interestingly, we found that magnetic helicity can destroy the chaos, which was not just due to inverse transfer of energy due to chaos as seen in two-dimensional turbulence. Magnetic helicity effectively increases the predictability of the system.
In other simulations, we also found a linear growth of error, as for the pure hydrodynamic case. However, this linear growth rate was separate for both fields, depending on the corresponding field's dissipation rate.
Since magnetic dissipation can be much smaller than the kinetic dissipation, the magnetic field has much greater predictability than the velocity field.
This suggests that the velocity details might vary greatly whilst the magnetic field is very similar.
Both of these effects suggest that galactic scale magnetic fields, if they have high magnetic helicity or Prandtl number, though turbulent, may have very long predictability times.


\section{Discussion and Conclusion}

In this review, we tried to highlight the contributions of many different researchers in advancing the field of knowledge of turbulence and specifically chaotic measures as a means to understanding it.
There are many different strands to this thought, but we pick out a few key ones below:

\begin{itemize}
    \item Turbulence is a multiscale phenomena both spatially and temporally. This is true even in chaos, where the different timescales interact to lead to a long term large length scale predictability co-existing with increasingly rapid error growth at the smallest scales.
    \item The Eulerian chaos and vortex stretching appear to be fundamentally linked, and may contribute to the changes seen in different dimensions.
    \item The Lyapunov spectrum may emerge as a practical measure for a complete description of the fluid flow.
    \item Chaotic measures, especially the maximum Lyapunov exponent, are more stable than spectral quantities, and show spatial features that are hard to discern from spectral quantities alone.
    \item By using chaotic measures, we can quantify simulations in a model free way that does not rely on a priori unknown spectral quantities, and thus improve computational efficiency.
    \item Chaotic measures can be used to detect phase transitions.
\end{itemize}

These recent discoveries are not only of theoretical importance, but of practical importance, since these measures are easily implementable in simulations of real world systems.
As well, we hope that this review will be useful for researchers exploring turbulence in other contexts, such as in spontaneous stochasticity \cite{eyink2020renormalization,bandak2024spontaneous}
and molecular-gas-dynamics simulations \cite{mcmullen2022navier},
where chaotic measures will also provide an alternative measure to spectral quantities to understand the difference between these and the Navier-Stokes equations.

\begin{acknowledgments}
RH acknowledges partial funding by the University of Oslo,
UiO:Life Science through the convergence environment ITOM. AB acknowledges partial funding by STFC.

For the purpose of open access, the author has applied a Creative Commons
Attribution (CC BY) licence to any Author Accepted Manuscript version
arising from this submission.
\end{acknowledgments}

\nocite{*}

\bibliography{refs}

\end{document}